# Anatomy of the vertebral column lymphatic network in mice

Laurent Jacob[1], Ligia Boisserand[2], Juliette Pestel[1], Salli Antila[3], Jean-Mickael Thomas[4], Marie-Stéphane Aigrot[1], Thomas Mathivet[5], Seyoung Lee[2], Kari Alitalo[3], Nicolas Renier[1], Anne Eichmann[5,6], Jean-Leon Thomas[1,2#]

[1]Université Pierre et Marie Curie Paris 06 UMRS1127, Sorbonne Université, Institut du Cerveau et de la Moelle Epinière, Paris, France

[2]Department of Neurology, Yale University School of Medicine, New Haven, CT, 06511, USA

[3]Wihuri Research Institute and Translational Cancer Biology Program, Faculty of Medicine, University of Helsinki, Helsinki, Finland

[4]Ecole Nationale Supérieure d'Art de la Villa Arson, 06100 Nice, France

[5]INSERM U970, Paris Cardiovascular Research Center, 56 Rue Leblanc, 75015 Paris, France

[6]Cardiovascular Research Center and the Department of Cellular and Molecular Physiology, Yale University School of Medicine, New Haven, Connecticut 06510-3221, USA

# corresponding author: jean-leon.thomas@yale.edu

**Abstract (158 words)**

Cranial lymphatic vessels (LVs) are involved in transport of fluids, macromolecules and CNS immune responses. Little information about spinal LVs is available, because these delicate structures are embedded within vertebral tissues and difficult to visualize using traditional histology. Here we reveal an extended vertebral column LV network using three-dimensional imaging of decalcified iDISCO-clarified spine segments. Spinal LVs are metameric circuits exiting along spinal nerve roots and connecting to lymph nodes and the thoracic duct. They navigate in the epidural space and the dura mater around the spinal cord, and associate with leukocytes, peripheral dorsal root and sympathetic ganglia. Spinal LVs are VEGF-C-dependent and remodel extensively after spinal cord injury. They constitute an extension to cranial circuits for meningeal fluids, but also a route for perineural fluids and a link with peripheral immune and nervous circuits. Vertebral column LVs may be potential targets to improve the maintenance and repair of spinal tissues as well as gatekeepers of CNS immunity.

**Introduction**

The lymphatic vasculature controls fluid homeostasis, macromolecular clearance and immune responses in peripheral tissues (Alitalo, 2011; Petrova and Koh, 2018). The brain was long considered to lack lymphatic vasculature, which has raised questions about how interstitial cerebral fluid (ISF) is cleared of waste products (Louveau *et al.*, 2017; Abbott *et al.*, 2018) and how immune surveillance of the brain is maintained (Ransohoff and Engelhardt, 2012; Engelhardt *et al.*, 2016; Filiano, Gadani and Kipnis, 2017). ISF is formed by water and small solutes that are exchanged through the capillary walls between the blood vessels and the brain. It has a similar composition to the cerebrospinal fluid (CSF) which drains the brain ventricles and meninges and is mainly produced in the choroid plexus (Zappaterra and Lehtinen, 2012). The CSF has been described to dynamically exchange with ISF along glial "lymphatic" (glymphatic) non-vascular periarterial routes, without crossing the endothelial cell layer, and subsequently to be cleared from the brain into the subarachnoid space (Iliff, Goldman and Nedergaard, 2015; Engelhardt *et al.*, 2016). The CSF outflow system involves specific extracranial lymphatic vasculature beds (Pollay, 2010; Proulx *et al.*, 2017). In addition, cranial meningeal LVs (mLVs) have been recently identified and were suggested to drain CSF into deep cervical lymph nodes (dcLNs) (Aspelund *et al.*, 2015; Louveau *et al.*, 2015; Antila *et al.*, 2017). In mice, cranial mLVs are mainly aligned alongside large dural venous sinuses, meningeal arteries and cranial nerves. Along the sagittal suture, the cranial lymphatic vasculature is rather discontinuous with valveless and small-diameter LVs, while it forms a larger network with valves toward the lateral and basal aspects of the skull (Aspelund *et al.*, 2015; Louveau *et al.*, 2015; Antila *et al.*, 2017). Meningeal lymphatic vasculature also exists in the skull of primates, including common marmoset monkeys and humans (Absinta *et al.*, 2017; Antila *et al.*, 2017).

VEGF-C expression in vascular smooth muscle cells and VEGFR3 in lymphatic endothelial cells (LECs) are essential for the development of cranial mLVs (Antila *et al.*, 2017; Blanchette and Daneman, 2017). The meningeal lymphatic vasculature develops later than the rest of the lymphatic network, first appearing at birth in the basal parts of the skull, then expanding during the neonatal period along dural blood vessels whose vascular smooth muscle cells supply the VEGF-C (Antila *et al.*, 2017). Classical immuno-histology on whole-mount preparations or cryo-sections showed that, during the first weeks after birth, LVs also developed a large network closely attached to the vertebral column (Antila *et al.*, 2017). Vertebral LVs were observed in intervertebral spaces, ventrally and dorsally to the midline, as well as along spinal nerve roots, laterally and together with blood vessels. In the cervical spine, vertebral LVs appeared directly connected with cranial mLVs located dorsally around the cisterna magna and ventrally around the foramen magnum. These observations suggested that cranial mLVs might extend caudally into the spinal lymphatic vascular system and connect from there to the peripheral lymphatic system. To test this model, we decided to produce a three-dimensional (3D)-map of the vertebral lymphatic system that respected structural interactions between the CNS and meninges, the surrounding bone and mesenchymal environment and the neighboring peripheral nervous system (PNS). This required us to preserve the overall bone structures around the CNS while simultaneously accessing and labeling the LVs of meninges contained within the protective layers of muscular and skeletal tissues.

To do so, we used the iDISCO (immunolabeling-enabled three-dimensional imaging of solvent-cleared organs) technique which enables volume imaging of immunolabeled structures in complex tissues, without disrupting tissue architecture (Renier *et al.*, 2014, 2016). Imaging of iDISCO treated vertebral segments with a light sheet fluorescent microscope (LSFM) revealed an extensive lymphatic vasculature inside the vertebral canal.

**Results**

**Lymphatic vasculature pattern in the thoracic spine**

To label vascular, immune and neural cell compartments within the intact vertebral column, we decalcified segments of 2-4 vertebrae dissected together with the surrounding muscle tissue along the rostro-caudal axis of the vertebral column. iDISCO tissue clearing and immunolabeling followed by light-sheet fluorescence microscope (LSFM) was used for 3D-reconstruction of the spinal LV network.

Because we previously described LVs mainly in the cervical vertebrae (Antila *et al.*, 2017), the iDISCO protocol was first applied to the thoracic spine, with the goal to characterize the three-dimensional extension of LVs in the vertebral column (**Fig. 1 A**). **Fig. 1 B** illustrates a lateral view of Alizarin red staining of bones within a cleared spinal column segment to reveal the vertebrae, intervertebral spaces and ligamentum flavum. **Fig. 1 C** shows a schematic latero-frontal perspective view of a thoracic vertebral segment. Lymphatic endothelial cells (LECs) were labeled using polyclonal antibodies against two well-established LEC markers, the LYVE-1 cell surface receptor and the nuclear Prox-1 transcription factor (Wigle and Oliver, 1999; Jackson *et al.*, 2001; Johnson *et al.*, 2008). Prox-1 labeled LYVE-1 positive LECs and LYVE-1 negative cells within the spinal cord that were previously identified as oligodendroglial cells (Kaser-Eichberger *et al.*, 2016) (**Fig. S1, A and B**). LYVE-1 labeled Prox1 positive LECs and Prox1-negative myeloid cells, as previously reported (Kaser-Eichberger *et al.*, 2016) (**Fig. S1, A and B**). LYVE-1 positive LECs within the vertebral column were negative for the blood vessel marker Podocalyxin (Testa *et al.*, 2009) (**Fig. S1, C and H**).

Despite labeling of some non-LECs, both markers clearly revealed a dense lymphatic network that was present between vertebrae and appeared confined to the intervertebral spaces (**Fig. 1, D and E and movies 1 and 2**). A few longitudinal vessels linked adjacent intervertebral lymphatic circuits together along the spinal cord (**Fig. 1, E and F**). Each vertebral LV was also connected to the peripheral lymphatic system surrounding the vertebrae, dorsally through the ligamentum flavum, dorsolaterally along the dorsal facet joint and ventrolaterally through the intervertebral foramen along ventral nerve roots and dorsal root ganglia (DRGs) (**Fig. 1, E and F**). The overall vertebral lymphatic circuitry was surprisingly dense compared to the previously described lymphatic vasculature of skull meninges (Aspelund *et al.*, 2015; Louveau *et al.*, 2015; Absinta *et al.*, 2017; Antila *et al.*, 2017).

We next used Imaris-3D image analysis software to illustrate the anatomy of meningeal lymphatic circuits within each vertebra and their relationship with the extra-vertebral LVs. Analyses were performed from global image acquisitions of the thoracic spine, such as the one on display in **movie 2** which shows a succession of vertebral lymphatic units along the rostro-caudal axis. Images were then segmented to generate a color-coded map of LV circuits. In **Fig. 1, G and H** and **movies 3** and **4** each color defines the Prox-1$^+$ vascular pattern of one vertebra along three successive thoracic vertebrae (red, blue, green) as well as the peripheral lymphatic vasculature (white). This identifies a metameric organization of spine LVs.

**Vertebral unit lymphatic architecture**

We next mapped the LV network in the vertebral canal from the dorsal to the ventral part of a vertebra. **Movie 2** and the corresponding view in **Fig. 2, A and B** show LVs around one segment of the thoracic spinal cord, and areas where higher magnification views were taken. Dorsally, semicircular lymphatic vessels navigate around the spinal cord, at the ventral border of the ligamentum flavum (**Fig. 2 C**). At the dorsal midline, these vessels contact lymphatic branches entering the epidural space from the overlying dense peripheral lymphatic vasculature in the intervertebral space between two spinous processes (**Fig. 2 D**). An inner and less complete LV circle lines the dura mater, hence two layers of LVs border the epidural space (**Fig. 2 D**). Laterally, at the level of the transverse facet joints, semicircular vessels including peripheral lymphatic vessels from the dorsal plexus converge toward a lymphatic circle (blue arrow in **Fig. 2 C**). From this point, lymphatic vessels redistribute either radially via the periphery, or ventrally toward the emergence of the ventral spinal nerve root (red double arrow in **Fig. 2 C)**. **Movie 2** allows to follow the route of peripheral lymphatic vessels from their entry at dorsal midline to their lateral exit from the vertebral canal. At the intervertebral foramen, dorsal root ganglia (DRGs) are covered by LVs that converge from the ventral and dorso-lateral circuits at their proximal and distal end, respectively (arrowheads in **Fig. 2 E**). In addition to these two circuits, a few longitudinal connecting vessels link vertebral lymphatic units together (**Fig. 2 F)**. Ventrally, a second circuit of semicircular lymphatic vessels converges toward the ventral spinal nerve root exit, while no lymphatic vessels are observed at the ventral midline (**Fig. 2 G)**.

We observed a similar LV pattern in other thoracic segments (n=5), allowing us to generate a schematic representation of these different compartments of the vertebral lymphatic circuitry with a specific color code for intervertebral circuits (red) and vertebral branches of the peripheral LV (blue) (**Fig. 2 H**).

**Epidural and dural vertebral lymphatic circuits are surrounded by myeloid cells**

To obtain 3D-resolution of LV localization in the spinal canal and meninges, Prox-1-labeled vertebral volumetric images were used to generate segmented images of membranes and the epidural space around the spinal cord. We enhanced the brightness of successive single image slices of Prox-1 labeled cervical and thoracic vertebrae to identify meninges, the epidural space and ligamentum flavum (**Fig. 3A**). We then manually delineated and colored the layers of meninges, dura mater or epidural space including the Prox1 label. **Fig. 3 A** shows one image slice with a color-code for meningeal layers (purple area) and the epidural space (green area). After 3D-reconstruction of a stack of similarly processed image slices, we generated color-coded layer masks for the arachnoid and dura mater together (purple area in **Fig. 3 B**), or the dura mater and the epidural space together (green area in **Fig. 3C**). The overlay of both masks revealed that Prox-1[+] LVs mainly localized in the epidural space (green), while the underlying dura mater layer (white) includes ventral LVs around DRGs and a few LVs on each side of the ventral and dorsal midline (**Fig. 3D** and **movie 5**). As shown on a lateral view (**Fig. 3E)**, dura mater LVs localized most extensively at bilateral DRGs. Interestingly, connecting vessels between two successive vertebrae (red arrows in **Fig. 3E**) navigate in the epidural space and appear to join LVs of the dura mater close to the DRGs (white arrows in **Fig. 3E**), suggesting a possible confluence of peripheral lymph and CSF at this level.

Vertebral LVs moreover appeared to be a privileged environment for immune myeloid cells, as was previously reported for skull lymphatics (Louveau *et al.*, 2015). Pre-cleared segments of the vertebral column co-labeled with antibodies against LYVE-1 and the common leucocyte antigen CD45 show that CD45$^+$ myeloid cells inside the vertebral canal are concentrated around LYVE-1$^+$ vertebral LVs (**Fig. 4, A-D**).

**LV patterning varies between cervical, thoracic and lumbar spine**

We next examined the lymphatic vasculature at cervical and lumbar levels of the vertebral column. A first approach by confocal imaging of wholemount preparations of cervical and lumbar vertebrae showed differences in the pattern of cervical and lumbar LVs. More specifically, extra-vertebral LVs on the dorsal aspect of the spine were more abundant in lumbar compared to cervical vertebrae (**Fig. S2, A-D**). The variability of LV patterning along the vertebral column was analyzed in further detail by volume imaging.

In the cervical region (**Fig. 5 A**), the dorsal extravertebral lymphatic plexus is reduced in size (green in **Fig. 5 B**) and intravertebral LVs exit ventrally and bilaterally through the intervertebral foramen to connect to cervical lymph nodes **(Fig. S2, E and F)**. The short interspace between cervical vertebrae is associated with a lack of longitudinal intervertebral vessels and a direct lymphatic vessel connection from the ventral root to the lateral lymphatic circle of the neighboring vertebra. Thoracic vertebral LVs, as described above, are defined by a large dorsal extravertebral plexus (**Fig. 5 C**) and a direct connection from ventrolateral spinal roots to the thoracic lymphatic duct **(Fig. 5 D)**. The thoracic and lumbar regions display similar extension and pattern of extravertebral and intravertebral LVs (**Fig. 5, C and E**). In lumbar vertebrae, the ventro-lateral circuits that exit on each side of the vertebral canal connect to lymph nodes. As shown in green in **Fig. 5 F,** lymphatic vessels circumvent the ventral body of the lumbar vertebra, converge on the ventral midline and split into two branches running toward the pair of aortic lumbar lymph nodes. Therefore, the vertebral LV architecture is conserved along the vertebral column, but the extension of extravertebral and intravertebral vessels around the spinal cord and their connection to the peripheral lymphatic system differs between the cervical, thoracic and lumbar vertebral levels.

**Vertebral lymphatic vasculature contacts the sympathetic nervous system**

We found that LVs covering the DRG dura mater (**Fig. 2F**) extend a collateral branch, bilaterally along the spine, that contacts paravertebral Prox1$^{low}$ ganglia (**Fig. 6, A and B**). Prox1 is known to be expressed in the sympathetic neuronal lineage (Holzmann, Hennchen and Rohrer, 2015) and double-labeling with antibodies against Prox-1 and Tyrosine Hydroxylase (TH), a specific marker of adrenergic nerves and ganglia, confirmed that specific branches emerging from vertebral LVs connect to TH$^+$/Prox1$^{low}$ sympathetic ganglia (**Fig. 6 C**) and that lymphatic vessels double stained with Prox-1 and LYVE-1 contact one sympathetic ganglion per spinal level (**Fig. 6 D-F**). Complementary analyses by high resolution confocal imaging on vertebral column cryosections indicated that the close spatial relationship between LV and sympathetic ganglia (**Fig. 6 G, H**) remained at the surface of the ganglion cortical layer (**Fig. 6I**). These data reveal a hitherto unknown anatomical interaction between the autonomous nervous system and lymphatic vessels derived from spinal meningeal LVs.

**Vertebral LVs respond to VEGF-C and spinal cord injury**

We next tested if vertebral LVs responded to the major lymphangiogenic growth factor VEGF-C. Adult mice (2months of age) were injected with AAV-VEGF-C or -empty control (n=4) into the cisterna magna, followed by analysis one month later (**Fig. 7 A**). **Fig. 7 B-E** and **movie 8** show Prox-1 staining of cervical vertebrae. Compared to control injected mice (**Fig. 7, B and D**), VEGF-C injected mice showed a strongly expanded LV network, in particular of dorsolateral lymphatic rings in the intervertebral disk (**Fig. 7, C and E** and **movie 6**).

The capacity of adult vertebral LVs to grow in response to growth factors suggested that this circuitry may respond to pathological conditions affecting the spinal cord and more broadly the CNS. Based on the prediction that a spinal cord trauma generates tissue destruction, accumulation of cell debris, and thus requires from the LV a supplement of tissue clearance and immune survey, we applied a direct injury at the thoraco-lumbar level, using a needle to inject sodium chloride (2μl) into the spinal cord (**Fig. 7F**). Within a week after the surgery, a robust lymphangiogenesis was observed in extravertebral and intravertebral lymphatic vasculature in the injured mice (**Fig. 7, G-N** and **movie 7)**.

**Discussion**

We here used the i-DISCO protocol combined with LSFM to generate a map of thoracic vertebral column LVs (**Fig. 8**). We reveal an extensive and complex lymphatic vasculature in the spinal column, surprisingly dense in comparison to the one that covers the cranial dura mater. Previous literature has reported the presence of LVs on whole mount preparations of vertebral dura mater in monkeys (Miura, Kato and von Lüdinghausen, 1998) and on sections of intact and injured vertebral tissue in humans (Kashima, Dongre and Athanasou, 2011). These studies identified elements of lymphatic vasculature in the spine, such as the dura mater and epidural lymphatic circuits and the absence of LVs in vertebral bones and intervertebral disks in the intact spine. We here extend these findings by 3D-views of lymphatic vasculature organization and reveal interactions with surrounding tissues along the spine (**Fig. 8**).

Each vertebra is drained by semicircular dorsal and ventral vessels, which exit the vertebral column at intervertebral foramen. Vertebral LVs extend along spinal nerve roots to reach either lymph nodes in the cervical and lumbar regions, or the thoracic lymphatic duct in the thoracic region. The vertebral lymphatic network is thus organized as a metameric network of peripheral LV-connected vertebral lymphatic units that are interconnected by a few thin longitudinal vessels. The absence of large longitudinal dorsal or ventral LVs suggests that the main flow of vertebral lymph is not descending as a continuous stream from the cisterna magna along the vertebral column axis, but is rather drained at the level of each vertebra. In addition to dura mater lymphatic circuits, the vertebral lymphatic vasculature includes an extensive network of epidural vessels that are located in the intervertebral tissue and beneath the ligamentum flavum and .appear to drain the non-CNS peripheral lymph of the vertebral column.

Pioneer studies in the late 19$^{th}$ and early 20$^{th}$ centuries (Schwalbe, 1869); (Quincke H, 1872); (Orr and Rows, 1907) and later works of Ivanow (Ivanow, G., 1927) and Brierley and Fields (Brierley and Field, 1948) had investigated the flow of the lymph stream along the spine and in the spinal roots of the cord as well as lymphatic pathologies and infectious agent propagation in the spine, providing 'a body of evidence that lymphatics play a part probably not subordinate to that of the blood vessels as channels by which infectious agents (toxins, polyomyelitis, tetanus) are conveyed to the cord and distributed within it (Bruce and Dawson, 1911). These predictions find support in the present imaging of extended vertebral lymphatic circuits that contact both peripheral lymph and CSF.

It is striking to note that there is a regional variation in the LV size, which is inversely correlated to the volume of CSF, with large cerebral ventricular volumes associated with a discrete network of cranial mLVs and a small vertebral ependymal volume correlated with a large vertebral lymphatic vasculature. One possible explanation is that vertebral LVs strongly contribute to the reuptake of the CSF that is continuously produced by the ventricular choroid plexus and circulates top-to-bottom along the spinal cord. This model is supported by presence of dura mater LVs in the spine and their dense location around spinal nerve root exits. A corollary of this hypothesis is that glymphatic drainage of interstitial fluids may be faster in the spinal cord than in the brain. A second and likely possibility is that the largest part of vertebral LV drains peripheral lymph, as suggested by the large network of epidural vessels that extends into the vertebral canal from the peripheral lymphatic system.

Vertebral LVs localize mainly at the level of intervertebral ligaments or joints, much like cranial lymphatics that navigate in skull commissures alongside blood vasculature and spinal nerves (Antila *et al.*, 2017). They also circulate through the fat tissue filling the epidural space intercalated between the dural mater and intervertebral ligaments. We and others find that vertebral lymphatic vessels avoid bone tissues (Edwards *et al.*, 2008). Interestingly, the presence of LVs inside bone is observed in patients with vanishing bone disease (also called Gorham Stout disease GSD) (Hominick *et al.*, 2018). GSD is a sporadic disease characterized by the presence of lymphatic vessels in bone and progressive bone loss. In severe cases, the disease progresses until entire bones are lost and replaced by fibrous tissue. GSD can affect any bone in the body, but it most frequently affects the ribs and vertebrae, with poor prognosis (Lala *et al.*, 2013; Dellinger, Garg and Olsen, 2014). In mice, transgenic VEGF-C overexpression in bone under the control of the Osx promoter induces lymphatic invasion into bone and osteoclast-mediated bone loss, mimicking essential aspects of GSD (Hominick *et al.*, 2018). Mechanisms preventing LV formation in bone are currently unknown.

In contrast to skull commissures between skull cap bones, which are few, narrow and fixed, the vertebral disks, joints and ligaments between vertebral bones are numerous, large and mobile. They sustain the integrity and flexibility of the spine, which is predicted to require extensive interstitial fluid drainage. The large network of vertebral epidural LV appears to be exquisitely adapted to this extensive drainage of non-neural peripheral tissues in the spine and to provide each vertebra with its proper clearance system. It is predictable that defective spine LVs will alter vertebral and intervertebral tissue maintenance, leading to spine orthopedic pathologies.

The spine LV circuitry includes epidural vessels in the peripheral lymphatic system that likely drain intervertebral non-neural tissues, especially the epidural fat tissue of the spinal cord, and dura mater vessels that may drain CNS tissue fluids at possible hot-spots close and around the DRGs as well as at more restricted positions on each side of the dorsal and ventral midline. These two circuits appear to be linked, since a part of especially LVs connecting vertebrae together, circulate through the epidural space before entering the dura mater around the DRGs and exiting the spinal canal. It remains to be determined whether both circuits converge to the same or different lymph nodes or lymphatic ducts.

The proximity of two distinct epidural and dura mater LV circuits in the spine raises questions about the protection of the privileged immune status of the CNS. Like in skull (Louveau *et al.*, 2015), a specific interaction between LVs and CD45 leucocytes is observed along the spine (**Fig. 3**). The spinal cord lymphatic vasculature thus appears as a potentially important immune surveillance interface between the CNS and peripheral tissues. The cervical and lumbar regions directly drain into cervical and lumbar lymph nodes, respectively, which suggests that the peri-lymphatic dendritic immune cells may rapidly transfer to lymph nodes and initiate lymphocyte activation against specific pathogens or antigens. On the other hand, the promiscuity of epidural and dura mater LV in the spine may predispose to the propagation of peripheral infections through the vertebral canal toward neural tissues, especially peripheral nervous ganglia which are in close contact with the spine LV (**Fig. 6**). For example, epidural LVs may provide entry for meningitis infection into spinal meninges. The contact zone between DRGs and LVs connected to the thoracic duct and lymph nodes appears as another potential gate for entry into the CNS for pathogens drained by the peripheral lymph. This possibility is in agreeement with the early detection of bovine scrapie protein in the mesenteric lymph nodes and DRGs of lemurs or cattles infected orally with the agent of bovine spongiform encephalopathy (BSE) (Bons *et al.*, 1999; Franz *et al.*, 2012).

The spine elicits a variety of diseases including infections caused by either a bacterial or a fungal infection transfered into the spine through the bloodstream (Darouiche, 2006), acute spinal cord compressions resulting from trauma, cancer, epidural abscess, or epidural hematoma (Ropper and Ropper, 2017), and degenerative spine disorders, a common condition especially in the ageing Western population (Brinjikji *et al.*, 2015). Vertebral LVs are a potential target for these pathologies, as they are pipes to propagate infections and to drain excessive interstitial fluids. The vertebral column is also the commonest site for skeletal metastatic tumors, with breast, prostate and lung cancers being the most common primary sources; as many as 70% of cancer patients have spinal metastases, and up to 10% of cancer patients develop metastatic cord compression (Choi *et al.*, 2010). Since lymphatics may serve as conduits for primary tumor cells in metastatic spreading (Karaman and Detmar, 2014), specific interference in the vertebral lymphatic vasculature could reduce or prevent the spinal metastasis process. Alternatively, lymphatic vessels are the first barrier for antigen-presenting cells that are transported into the next lymph node to initiate an adaptive immune response (Card, Yu and Swartz, 2014). Facilitating the entry of immune cells into vertebral lymphatic vessels could thus also potentially improve the efficiency of immune checkpoint inhibitor treatments to trigger the adaptive immune response against spinal tumor cells.

We find that adult vertebral LVs rapidly expand in response to VEGF-C or tissue injury. Tissue injury causes inflammatory processes and calls for repair mechanisms. These processes require increased LV activity and thus lymphangiogenesis which is likely VEGF-C/VEGFR3 signaling dependent, as in chronic skin inflammation (Hagura *et al.*, 2014) and macrophage VEGF-C-driven (Ji, 2012). Spine LVs and VEGF-C may thus be new therapeutic targets to support and improve these attemps to protect and repair injured spine tissues.

Vertebral LVs never contact the spinal cord tissue, even upon VEGF-C overexpression or acute spine lesion. Like cranial LVs, vertebral LVs are thus unlikely to reuptake CSF circulating through perivascular spaces of the spinal cord and pia mater. In contrast, vertebral LVs are closely apposed around the somatic and autonomous nervous ganglia chains. The targeting of paravertebral sympathetic chains by specific vertebral LVs is another discovery resulting from the spine volume imaging. Although no lymphatic vascularization of sympathetic ganglia was observed, lymphatic vessels may provide molecular signals to the sympathetic neurons that control vascular tone of lymphatic ducts and cerebral arteries and arterioles. Previous observations also showed that adrenergic fibers connect to the thoracic lymphatic duct and also innervate the wall of lymph node arterioles (Villaro, Sesma and Vazquez, 1987; Mignini *et al.*, 2012). The crosstalk between spine LV and the sympathetic system is thus likely relevant for the regulation of peripheral lymph and glymphatic drainage and may coordinate them with the activity of brain and spine tissues. We speculate that a regulatory loop may link meningeal LV, sympathetic chain neurons and both CNS and peripheral fluid drainage.

To conclude, this study shows that the volume imaging technique allows the description of neurovascular systems by preserving the anatomy and the 3D-continuity of vascular and neural structures. In particular, we have revealed a new set of information on the anatomical organization and plasticity of the lymphatic vasculature along the spine. Our findings identify vertebral LVs as a potential targets for improving the maintenance and repair of vertebral tissues as well as a gatekeeper of CNS immunity.

**Acknowledgements**

This work was supported by Institut National de la Sante et de la Recherche Medicale (to J.-L.T.), Agence Nationale Recherche (ANR-17-CE14-0005-03 to J.-L.T. and A.C.E.), Federation pour la Recherche sur le Cerveau (FRC 2017 to J.L.T. and A.C.E), Carnot Maturation (to L.J.), Leducq Foundation Transatlantic Network of Excellence (ATTRACT to A.C.E.), NIH (R01EB016629-01 to J.-L.T., NHLBI 1R01HLI125811, NEI 1R01EY025979-01, P30 EY026878 to A.C.E.) and the Yale School of Medicine (J.L.T.). K.A. funding was provided by the Jane and Aatos Erkko Foundation, the European Research Council (European Union Horizon 2020 research and innovation program, grant 743155), the Wihuri Foundation, the Academy of Finland (Centre of Excellence Program 2014-2019, grants 271845 and 307366), and the Finnish Brain Foundation. We acknowledge the ICM-QUANT cellular imaging, ICM-histomics and PHENO-ICMICE platforms, Andrey Anisimov and the AAV Gene Transfer and Cell Therapy Core Facility of Biomedicum Helsinki, University of Helsinki, for providing us AAV-expression vectors.

**Figure legends**

**Fig 1. Segmental pattern of the vertebral lymphatic vasculature in the thoracic spine**

(A) Alcian blue/Alizarin red staining of the mouse vertebral column with boxes indicating position of images shown in Figs 1-4 (thoracic vertebrae) and 5 (cervical and lumbar vertebrae), spatial orientation (A: anterior, D: dorsal, L: lateral, V: ventral). (B, C) Alizarin red staining of a thoracic vertebral column segment (B) and corresponding latero-frontal perspective schematic view (C). Two successive vertebrae are delimited by red/blue dots. DM: dura mater, LF: ligamentum flavum, LM: leptomeninges (pia mater and arachnoid), SC: spinal cord, SN: spinal nerve, blue arrow: dorso-lateral facet joint, red arrowhead: ventral intervertebral disk, blue star: intervertebral foramen, red (B) or black (C) star: ventral vertebral body. (D) Dorsal view of LYVE-1 staining. Red and blue areas correspond to two successive vertebrae. Note LVs lining ligamentum flavum. (E, F) Dorsal (E) and lateral (F) views of the Prox-1 expression pattern. Red and blue areas correspond to two successive vertebrae. Salmon arrows: intervertebral LVs, red arrow: dorsal LVs, red star: vertebral ventral body. (G, H) Segmented images of the Prox-1 LV network (fronto-dorsal (G) and lateral (H) views) highlighting three successive vertebral LV units (red, blue, green).

**Fig 2. Organization of individual thoracic vertebral column lymphatic units**

(A) Frontal view of a cleared thoracic vertebra stained with an anti-Prox-1 antibody. Red star: vertebral ventral body, SC: spinal cord. (B) Magnification of red boxes referring to images in C-G. (C) Semicircular dorsal LVs (red arrow) surround the spinal cord, exit dorso-laterally (blue arrow) and also extend a latero-ventral connection (double red arrows) to the ventral nerve root (see double arrow in F). Note Prox1 positive cells in SC and perivertebral muscles (M), FJ: facet joint. (D) At the ventral face of the ligamentum flavum (LF) located between two spinous processes, dorsal LVs (blue arrows) enter the vertebral canal and join semicircular LVs (red arrows). Note circles of LVs bordering the upper side of the epidural space (ES). (E) Ventro-lateral LV circuitry around DRG (red arrowhead). Blue dotted-lines: spinal nerve roots, red dotted-lines: DRG. (F) Lateral view with intervertebral LVs (salmon arrows), red dotted-lines: DRG. (G) Two ventral branches arising on each side of the ventral midline (VM) and reaching the DRG (see single red arrows in F). (H). Schematic representation of a frontal view of a thoracic vertebral LV unit. Longitudinal connecting vessels between vertebral units are not represented. FJ: facet joint; LF: ligamentum flavum; ES: epidural space; DRG: dorsal root ganglia; VM: ventral midline; SC: spinal cord. Black letters referring to images in C-G.

**Fig 3. Epidural and dural lymphatic circuits of the spine**

(A) 2D-single frontal image slice (2µm thick) of the cervical vertebral column with enhanced brightness to reveal Prox-1-expressing nuclei and spinal tissues including spinal cord (SC), meninges including pia mater (P), arachnoid (A) and dura mater (D), the epidural space (ES) and the ligamentum flavum (LF). A color-coded segmentation of layers around the spinal cord shows the meninges in purple and the dura mater plus the epidural space in green. (B-D) 3D-reconstruction of frontal images of the cervical vertebral column with color-coded layers: the arachnoid and dura mater in purple (B); the dura mater and epidural space in green (C); combined layer marks showing the arachnoid in purple, the dura mater in white and the epidural space in green (D). A noticeable LV network fills the epidural space (green) while dura mater LVs (white) are mainly restricted to DRGs (white arrows) and few branches on each side of the dorsal and ventral midline. (E) 3D-reconstruction of lateral images of the thoracic vertebral column with color-coded layers illustrated in (D). Blue dotted-lines: bilateral DRGs; salmon arrows: intervertebral LVs; Red star: vertebral ventral body.

**Fig 4. Interactions of spinal LVs with immune cells**

Double labeling of cleared thoracic vertebral column segments with LYVE-1 (purple in A, C, D) and CD45 (green in A, B, D) to identify myeloid cells. (B-D) are magnifications of white box in (A). Merged pictures (D). Note that myeloid cells concentrate along LVs. White star: vertebral ventral body; SC: Spinal cord.

**Fig 5. Variations of LV patterns along the vertebral column**

Pattern of Prox-1$^+$ LVs in the cervical (A, B), thoracic (C, D) and lumbar (E, F) vertebral column. Left panels show frontal views, right panels show connection to peripheral lymph nodes (LN) and thoracic duct (TD). (A, C, E). Note fewer LVs in the dorsal plexus between intervertebral spinous processes of cervical and lumbar vertebrae compared to thoracic ones (blue arrows). Also note differences in ventral root exit circuits between regions (salmon arrows, A ,C, E). (B, D, F). LV ventral exit circuits (green) and deep cervical LNs (B), thoracic duct (D) or connections to aortic lumbar LNs (F). Red star: vertebral ventral body; SC: spinal cord; Ao: Aorta.

**Fig 6. Lymphatic vessel connections with sympathetic ganglia**

(A, B). Thoracic Prox-1$^+$ LVs contact paravertebral sympathetic ganglia (SG) (blue arrow),. White dotted-lines: DRG. (C) Prox-1 (white) and tyrosine hydroxylase (TH, red) double labeling shows a ventral LV branch contacting a paravertebral TH$^+$ sympathetic ganglion (blue arrow). (D-F). Prox-1 (white) / LYVE-1 (red) double-labeling of the LV-SG connection (blue arrow). (G-I). 2D-confocal images of cervical cryosections labeled with LYVE-1 (white), TH (red), and Dapi (Blue). White box: area magnified in (H) and (I). A LV contacts a TH$^+$ SG (blue arrow). Note a second ventral LV branch running along the SG, without entering its cortical layer (G-I, salmon arrow). This branch is also seen in panel (B) (salmon arrow). White dotted-lines: DRG; Red star: vertebral ventral body; SC: spinal cord.

**Fig 7. Spinal LVs expand in response to VEGF-C and injury**

(A-E). Response of vertebral LVs to AAV-control or AAV-VEGF-C delivery into the CSF. (A) 2-mo mice received Intra-cisterna magna injection of AAVs and were analyzed by 3D-iDISCO one month later. Frontal views (B, C) and lateral views (D, E) show that AAV-VEGF-C (C, E) induced lymphangiogenesis in the cervical vertebral column as compared to AAV-control (B, D). Red star: vertebral ventral body; SC: Spinal cord.

(F-J) Spinal cord (Sc) lesion. (F) Experimental procedure: 2ul NaCl were injected into the spinal cord at the indicated level and mice were analyzed one week later. Frontal views (G, H) and lateral views (I, J) of Prox-1$^+$ LVs. Note that SC lesion induces robust lymphangiogenesis but LVs do not enter lesioned SC tissue. Red star: vertebral ventral body; SC: Spinal cord.

**Fig 8. Schema of thoracic vertebral column LVs**

LVs around spinal cord (gray), in the epidura space (green), in the dura mater (purple), extra-vertebral LV ventral connexion with SG and TD (blue). Blue dots indicate connexion with extra-vertebral lymphatic networks. Black star: vertebral ventral body; DRG: dorsal root ganglia, FJ: facet joint, LF: ligamentum flavum, SC: Spinal cord; SG: sympathetic ganglia.

**Supplementary data:**

**Fig. S 1: LVs and blood vasculature in the thoracic vertebral column**

(A, B) Prox-1 (white) / LYVE-1 (red) double staining of a clarified thoracic vertebral column segment. Note overlap of Prox-1 and LYVE-1 labeling in LVs (arrowheads). Also note Prox-1-labeling of SC oligodendrocytes (A) and LYVE-1 staining of myeloid cells in meninges (arrows, B). (C-H) Double labeling of a clarified thoracic vertebral column segment with LYVE-1 (green) and Podocalyxin (purple) to identify blood vessels. F-H: magnifications of boxed areas in C-E. White star: vertebral ventral body; SC: Spinal cord.

**Fig. S 2 : LV patterning varies between cervical, thoracic and lumbar spine**

(A-D) Whole-mount immunostaining of *Prox1-eGFP* reporter mice showing lymphatic vasculature (green, gray). (A) Dorsal lymphatic vasculature around cisterna magna and vertebrae in cervical and thoracal regions. (B) Close-up of lymphatic vessels around cisterna magna in close association with the LVs around first cervical vertebra. (C) Side-view of the vertebral LVs connecting longitudinally with each other (marked with blue asterisks) and surrounding the spinal nerves (SN). (D) Close-up of vertebral lymphatic vessels in the dorsal aspect of the spine showing their localization mainly in intervertebral spaces. Blue arrowheads mark lymphatic valves. Data shown are representative *n*=3-6 per one staining. Scale bars: 1 mm (A), 200 µm (B, C), 400 µm (D).

(E, F) Prox-1-staining of a cervical vertebral segment shows connection between a LV surrounding the DRG and a dcLN (blue arrows). (F): higher magnification view of boxed area in (E). Red star: vertebral ventral body; SC: Spinal cord.

**Movie 1: Lymphatic vasculature of a thoracic vertebral segment labeled with LYVE-1 antibody**

**Movie 2: Lymphatic vasculature of a thoracic vertebral segment labeled with Prox-1 antibody**

**Movie 3: Time-lapse *z* series of the Prox-1-labeled lymphatic vasculature of a thoracic vertebral segment**

**Movie 4: 3D-image segmentation of Prox-1-labeled LVs of a thoracic vertebral segment**

**Movie 5: 3D-image segmentation of Prox-1-labeled LVs in meningeal and epidural layers around the cervical spinal cord.**

**Movie 6: Prox-1-labeled cervical LVs following AAV-VEGF-C administration**

**Movie 7: Prox-1-labeled lumbar LVs at one week after spinal cord injury**

**Materials and methods**

**Study approval**

All mouse experiments were approved by the Institutional Animal Care and Use Committee of ICM Brain and Spine Institute and Yale University.

**Mice**

C57BL/6 background mice were used for all the experiments. Littermate mice were used as controls. Mice were considered as adults at 2 to 3-months (mo) of age.

**Tissue preparation**

Mice were given a lethal dose of Sodium Pentobarbital (Euthasol Vet) and perfusion-fixed through the left ventricle with 10ml ice-cold PBS then 20ml 4% paraformaldehyde (PFA) in PBS after puncture of the right atrium. To dissect the spine, the skin was completely removed, all the organs were eliminated and the ribs were completely cut to keep the vertebra from the cervical part until the lumbar part with the whole spinal cord inside. All the surrounding tissue as muscles, aorta, ligaments were maintained around vertebral column. The spine was cut into pieces of about 0,5cm (1 to 3 vertebrae) corresponding to the cervical, thoracic and lumbar regions. The different spinal segments were immediately immersed in ice-cold 4% PFA, fixed overnight at +4°C, washed in PBS, and processed for staining.

**Samples pretreatment in methanol for iDISCO protocol**

We used the classical iDISCO protocol ((Renier *et al.*, 2014) http://www.idisco.info). In detail, fixed samples were dehydrated progressively in methanol/PBS, 20%, 40%, 60%, 80% and 100% for 1 hr each (all steps were done with agitation). They were then incubated overnight in a solution of Methanol 33%/dichloromethane 66% (DCM) (Sigma 270997-12X100ML). After 2 x 1h washes with methanol 100%, samples were then bleached with 5% H2O2 in methanol (1 vol 30% $H_2O_2$/ 5 vol methanol) at 4°C overnight. After bleaching, samples were rehydrated in methanol for 1 hr each, 80%, 60%, 40%, 20% and PBS. To clarify vertebral bone, we here added a decalcification step using Morse solution (1/1 tri-sodium citrate and 45% formic acid) during 30min at RT. Samples were then washed rapidly with PBS then incubated 2x 1h in PTx2 (PBS/0.2% Triton X-100). At this step they were processed for immunostaining.

**Immunolabelling iDISCO protocol**

Pretreated samples were incubated in PBS/0.2% Triton X-100/20% DMSO/0.3M glycine at 37°C for 24h, then blocked in PBS/0.2% Triton X-100/10% DMSO/6% Donkey Serum at 37°C for 24h. Samples were incubated in primary antibody dilutions in PTwH (PBS/0.2% Tween-20 with 10mg/ml heparin) /5% DMSO/3% Donkey Serum at 37°C for 6 days. Samples were washed five times in PTwH until next day, and then incubated in secondary antibody dilutions in PTwH/3% Donkey Serum at 37°C for 4 days. Samples were finally washed in PTwH five times until the next day before clearing and imaging.

**iDISCO tissue Clearing**

Immunolabeled samples were dehydrated progressively in methanol in PBS, 20%, 40%, 60%, 80% and 100% each for 1 hr. They were then incubated overnight in a solution of Methanol 33%/DCM 66% followed by incubation in 100% DCM for 2x 15 minutes to wash the methanol. Finally, samples were incubated in DiBenzyl Ether (DBE) (without shaking) until cleared (4 hr) and then stored in DBE at room temperature before imaging.

**Cryostat immunostaining**

For cryosections of the spinal canal, the fixed tissues underwent decalcification with 0.5 M EDTA, pH 7.4, at 4°C. When the bone was soft, samples were washed thoroughly with PBS and immersed in PBS containing 20% sucrose and 2% polyvinylpyrrolidone for 24 h at 4°C, embedded in OCT compound (Tissue-Tek), and frozen for storage at −80°C. 50–100-μm-thick sections were cut using a cryostat (Microm HM 550/CryoStar NX70; Thermo Fisher Scientific), air-dried, encircled with a pap pen, permeabilized with 0.3% PBS-TX, washed with PBS, and blocked in 5% donkey serum in PBS-TX at RT. After overnight primary antibody incubation at 4°C in the same solution, the sections were washed with PBS and incubated with the appropriate fluorophore-conjugated secondary antibodies diluted in 0.3% PBS-TX for 1–2 h at RT. After washes with PBS, the sections were mounted with Vectashield mounting medium (Vector Laboratories), sealed with Cytoseal 60, and imaged as soon as possible.

**Antibodies**

The following primary antibodies were used for iDISCO protocol of mouse tissues: Rat anti–mouse podocalyxin (1:2500 MAB1556; R&D Systems), goat anti–human PROX1 (1:1200, AF2727; R&D Systems), rabbit anti–mouse LYVE-1 (1:800;11-034, AngioBio), goat anti-mouse CD45 (1:2000, AF114; R&D Systems), rabbit anti–mouse Tyrosine Hydroxylase (1:1500, T9237-13, United States Biological).

The primary antibodies were detected with the corresponding Alexa Fluor -555, -568 or -647 conjugated secondary antibodies from Jackson ImmunoResearch at 1/1000 dilution in all different conditions.

**Light-Sheet and Confocal Imaging**

Cleared samples were imaged in transverse orientation with a light-sheet microscope (Ultramicroscope II, LaVision Biotec) equipped with a sCMOS camera (Andor Neo) and a 23/0.5 objective lens (MVPLAPO 23) equipped with a 6 mm working distance dipping cap. Version v144 of the Imspector Microscope controller software was used. The microscope chamber was filled with DBE. We used one laser configuration. The light sheet was generated by scanning a supercontinuum white light laser, filtered with the following excitation band pass filters: 560/30nm (Alexa Fluor -568 or- 555) or -617/83nm (Alexa Fluor-647). In this configuration we matched the xy width of the light sheet to the sample and NA. Samples were generally imaged under 45% laser power. We used the following emission filters: 595/40 for Alexa Fluor-568 or -555, and -680/30 for Alexa Fluor-647. Stacks were acquired using 2µm z steps and a 30ms exposure time per step. Pixel sizes ranged from 5.16 to 0.52 µm, with the zoom factor used for acquisition (8X). Mosaic acquisitions were done with different number of frame (X, Y) 2x2; 3x2 or 3x3 in function of the samples size.

Laser scanning confocal micrographs of the fluorescently labeled brain were acquired using a Leica TCS SP8 confocal microscope (air objectives 10× Plan-Apochromat with NA 0.45 and 25× Plan-Apochromat with NA 1.1) with multichannel scanning in-frame.

**Image Processing and Analysis**

For display purposes in the figures, a gamma correction of 1.47 was applied on the raw data obtained from the light-sheet microscope.

Images acquired with Imspector acquisition software in tif fomat was converted with Imaris File Converter to have IMS files. Mosaics were reconstructed with Imaris stitcher then Imaris software (Bitplane, http://www.bitplane.com/imaris/imaris) was used to: generate the orthogonal projections of data shown in all figures; perform area segmentation on a stack of image slices and apply a color code to selected lymphatic networks.

**AAV injection**

Intra cisterna magna (CM) injection was done into adult male C57BL/6J mice of 8–10 wk of age. A single dose of 2 µl ($10^9$ viral particles per µl) of AAV encoding VEGF-C (Anisimov *et al.*, 2009). Mice were anesthetized with isoflurane (induction 4%, maintenance 2%) and placed in a stereotactic apparatus (Stoelting). Injection into the CM was done using a Hamilton syringe with a 34-G needle and a flow rate of 0.5 µl/min. The needle tip was retracted 2 min after the injection. 100µl of 0.05 mg/kg of Buprecar solution (administered subcutaneously) was used to relieve postoperative pain. The AAVs of serotype 9 were produced by the AAV Gene Transfer and Cell Therapy Core Facility, HiLIFE, University of Helsinki (He *et al.*, 2005; Bry *et al.*, 2010; Calvo *et al.*, 2011).

**Spinal cord lesion**

Adult male C57BL/6J mice of 8–10 wk of age were used. Lesions were induced in the spinal cord by a stereotaxic injection of sterile 0.9% NaCl solution. Prior to the surgery, mice were anesthetized by intraperitoneal injection of ketamine (90mg/kg) and xylazine (20mg/kg) cocktail. Two longitudinal incisions into longissimus dorsi at each side of the vertebral column were performed, and the muscle tissue covering the column was moved to the side. Animals were placed in a stereotaxic frame, the 13th thoracic vertebra was fixed in between the bars designed for manipulations on mouse spinal cord (Stoelting,Wood Dale, IL), and intravertebral space was exposed by removing the connective tissue. An incision into dura mater was performed using a 30-gauge needle, and NaCl was injected using a glass micropipette attached via a connector to a Hamilton's syringe and mounted on a stereotaxic micromanipulator. Following injection, the muscle sheaths were sutured with 3/0 Monocryl, and the skin incision was closed with 4/0 silk. After 7 days post injection, the mice were perfused with 4% PFA; tissues were harvested and processed for iDISCO protocol as described above.

Blanchette, M. and Daneman, R. (2017) 'The amazing brain drain', *The Journal of Experimental Medicine*, 214(12), pp. 3469–3470. doi: 10.1084/jem.20172031.

Bons, N. *et al.* (1999) 'Natural and experimental oral infection of nonhuman primates by bovine spongiform encephalopathy agents', *Proceedings of the National Academy of Sciences*, 96(7), pp. 4046–4051. doi: 10.1073/pnas.96.7.4046.

Brierley, J. B. and Field, E. J. (1948) 'The connexions of the spinal sub-arachnoid space with the lymphatic system', *Journal of Anatomy*, 82(Pt 3), pp. 153–166.

Brinjikji, W. *et al.* (2015) 'Systematic literature review of imaging features of spinal degeneration in asymptomatic populations', *AJNR. American journal of neuroradiology*, 36(4), pp. 811–816. doi: 10.3174/ajnr.A4173.

Bruce, A. and Dawson, J. W. (1911) 'On the relations of the lymphatics of the spinal cord', *The Journal of Pathology and Bacteriology*, 15(2), pp. 169–178. doi: 10.1002/path.1700150204.

Bry, M. *et al.* (2010) 'Vascular endothelial growth factor-B acts as a coronary growth factor in transgenic rats without inducing angiogenesis, vascular leak, or inflammation', *Circulation*, 122(17), pp. 1725–1733. doi: 10.1161/CIRCULATIONAHA.110.957332.

Calvo, C.-F. *et al.* (2011) 'Vascular endothelial growth factor receptor 3 directly regulates murine neurogenesis', *Genes & Development*, 25(8), pp. 831–844. doi: 10.1101/gad.615311.

Card, C. M., Yu, S. S. and Swartz, M. A. (2014) 'Emerging roles of lymphatic endothelium in regulating adaptive immunity', *The Journal of Clinical Investigation*, 124(3), pp. 943–952. doi: 10.1172/JCI73316.

Choi, D. *et al.* (2010) 'Review of metastatic spine tumour classification and indications for surgery: the consensus statement of the Global Spine Tumour Study Group.', *European spine journal : official publication of the European Spine Society, the European Spinal Deformity Society, and the European Section of the Cervical Spine Research Society, European Spine Journal*, 19, 19(2, 2), pp. 215, 215–222. doi: 10.1007/s00586-009-1252-x, 10.1007/s00586-009-1252-x.

Darouiche, R. O. (2006) 'Spinal epidural abscess', *The New England Journal of Medicine*, 355(19), pp. 2012–2020. doi: 10.1056/NEJMra055111.

Dellinger, M. T., Garg, N. and Olsen, B. R. (2014) 'Viewpoints on vessels and vanishing bones in Gorham-Stout disease', *Bone*, 63, pp. 47–52. doi: 10.1016/j.bone.2014.02.011.

Edwards, J. R. *et al.* (2008) 'Lymphatics and bone', *Human Pathology*, 39(1), pp. 49–55. doi: 10.1016/j.humpath.2007.04.022.

Engelhardt, B. *et al.* (2016) 'Vascular, glial, and lymphatic immune gateways of the central nervous system', *Acta Neuropathologica*, 132, pp. 317–338. doi: 10.1007/s00401-016-1606-5.
23

Fig 1

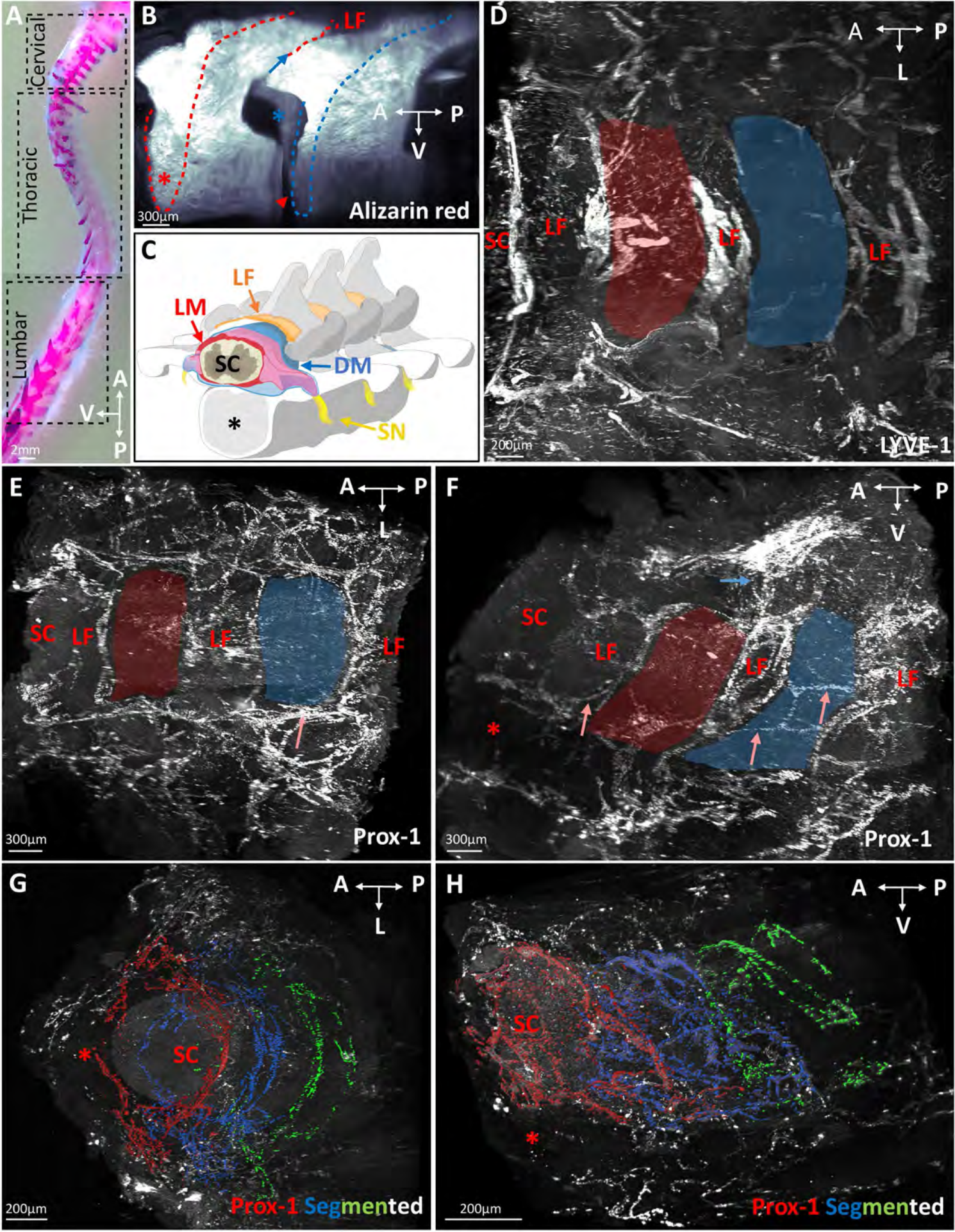

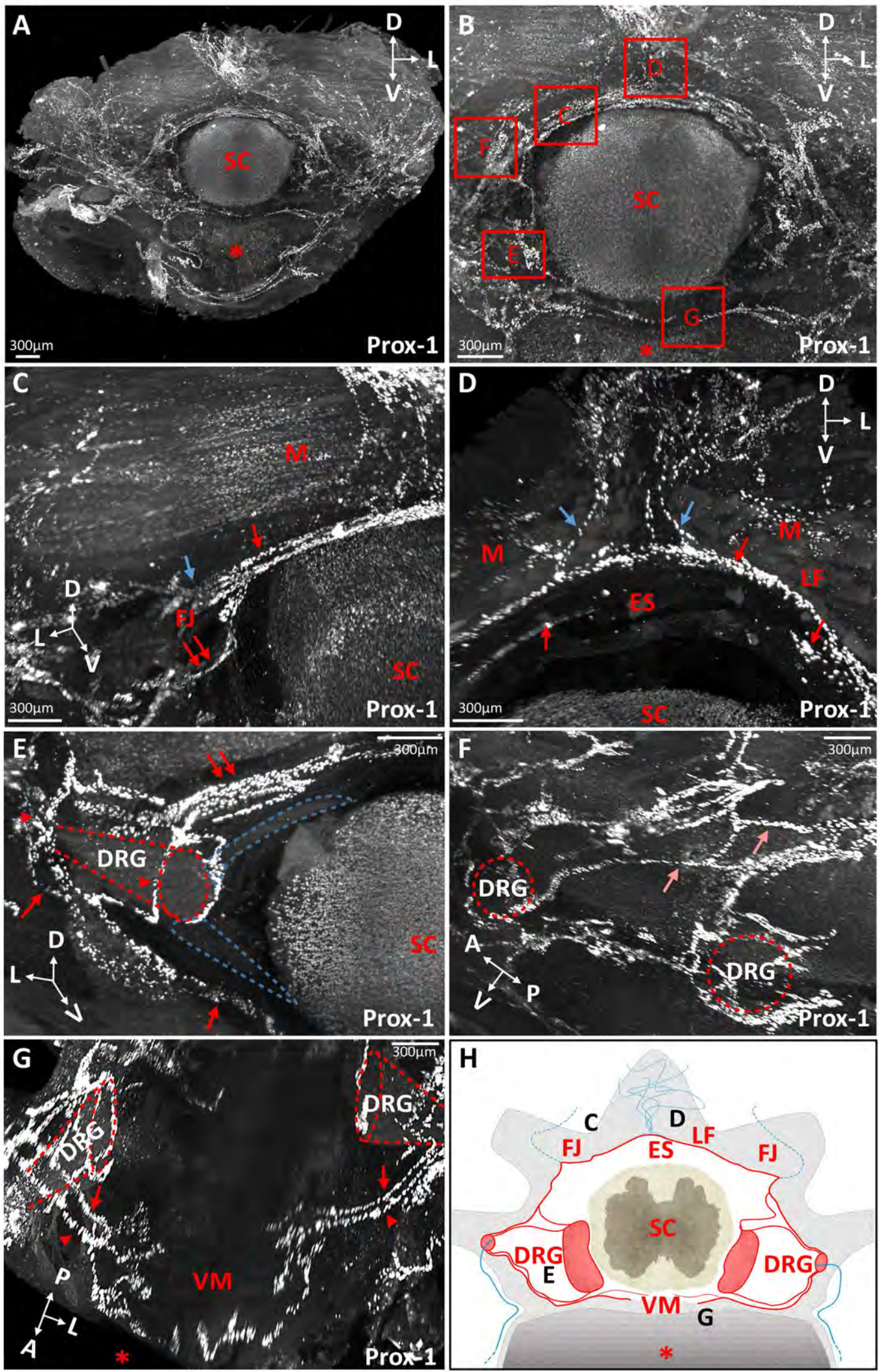

Fig 2

Fig 3

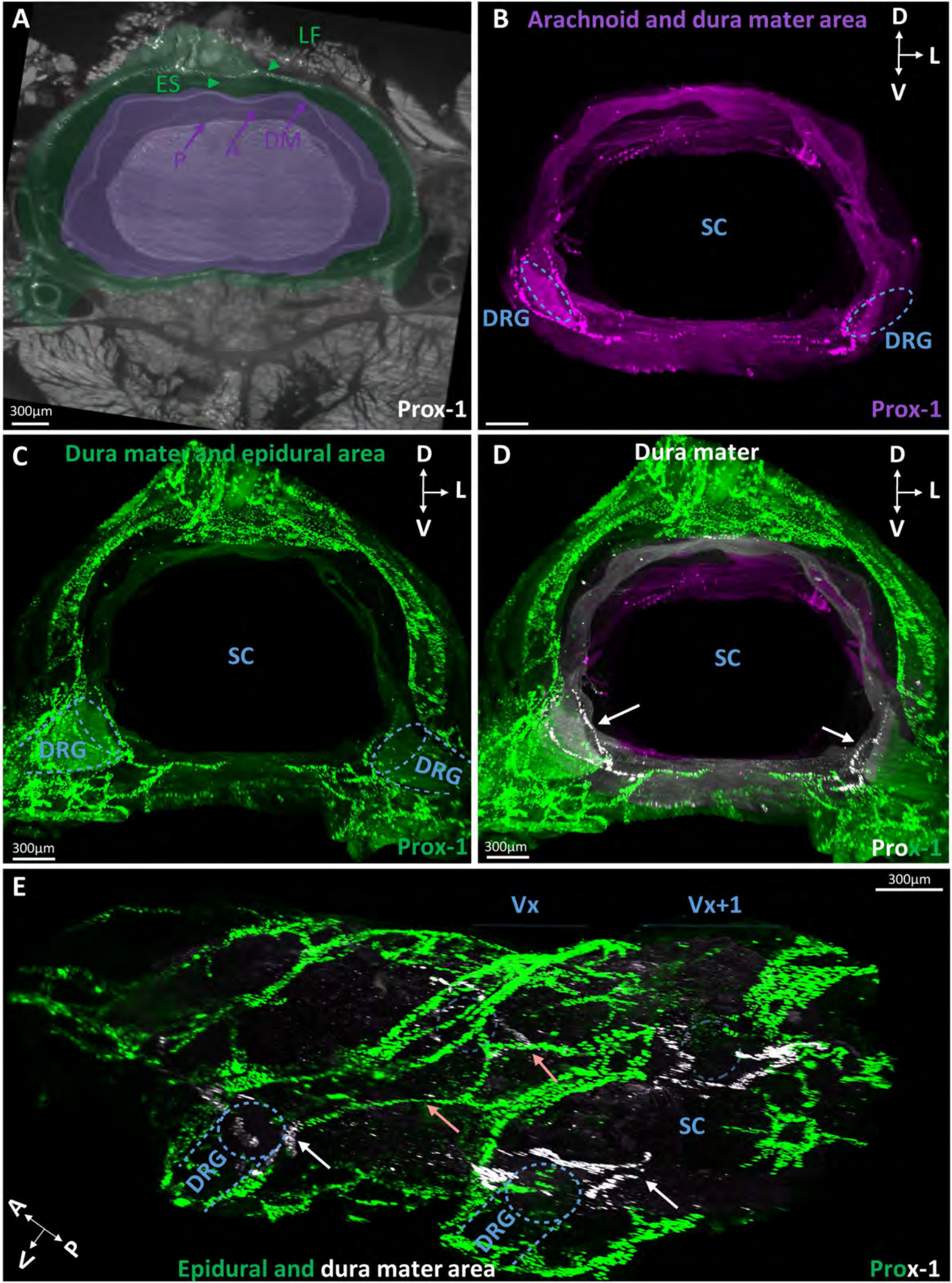

Fig 4

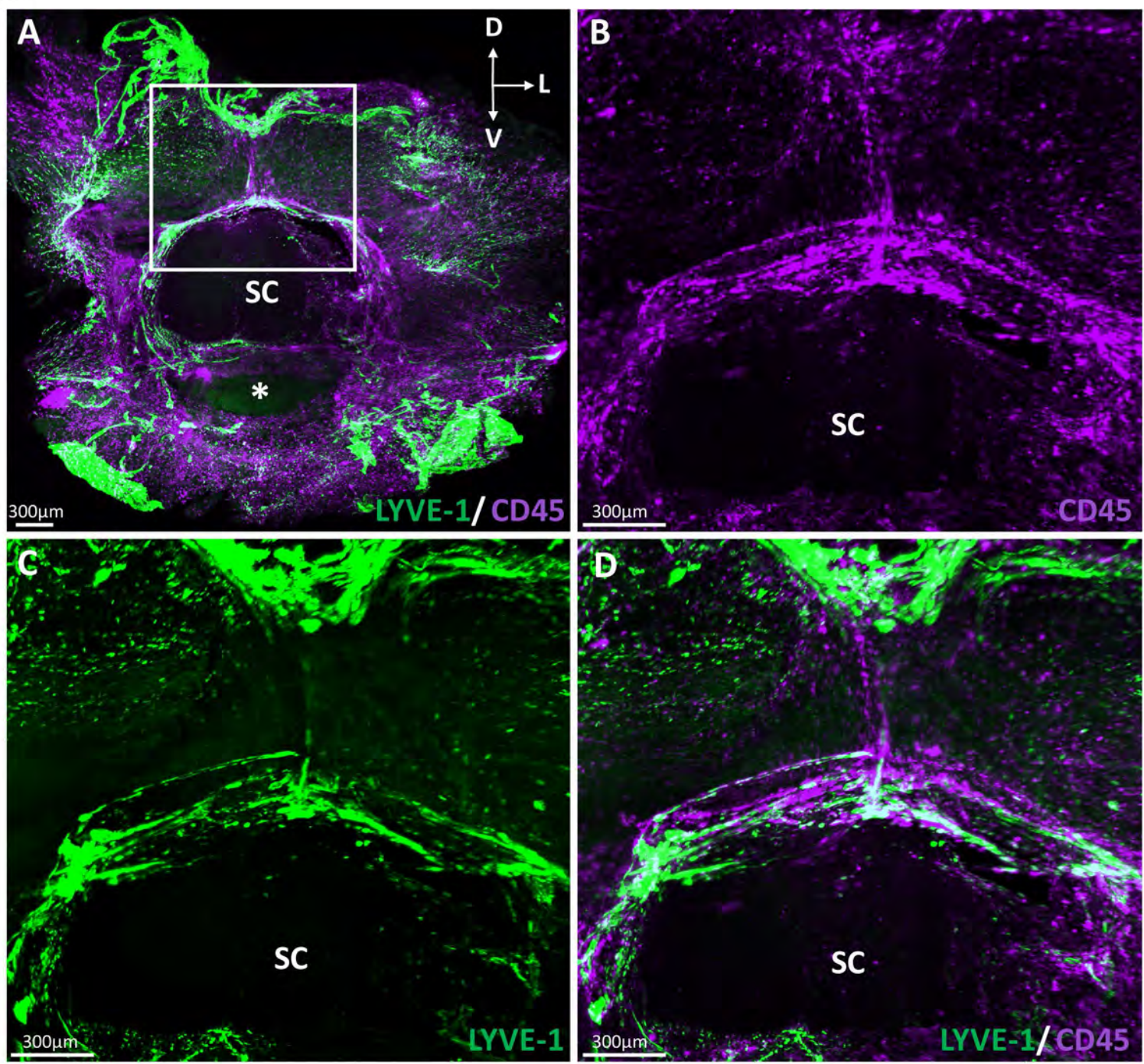

Fig 6

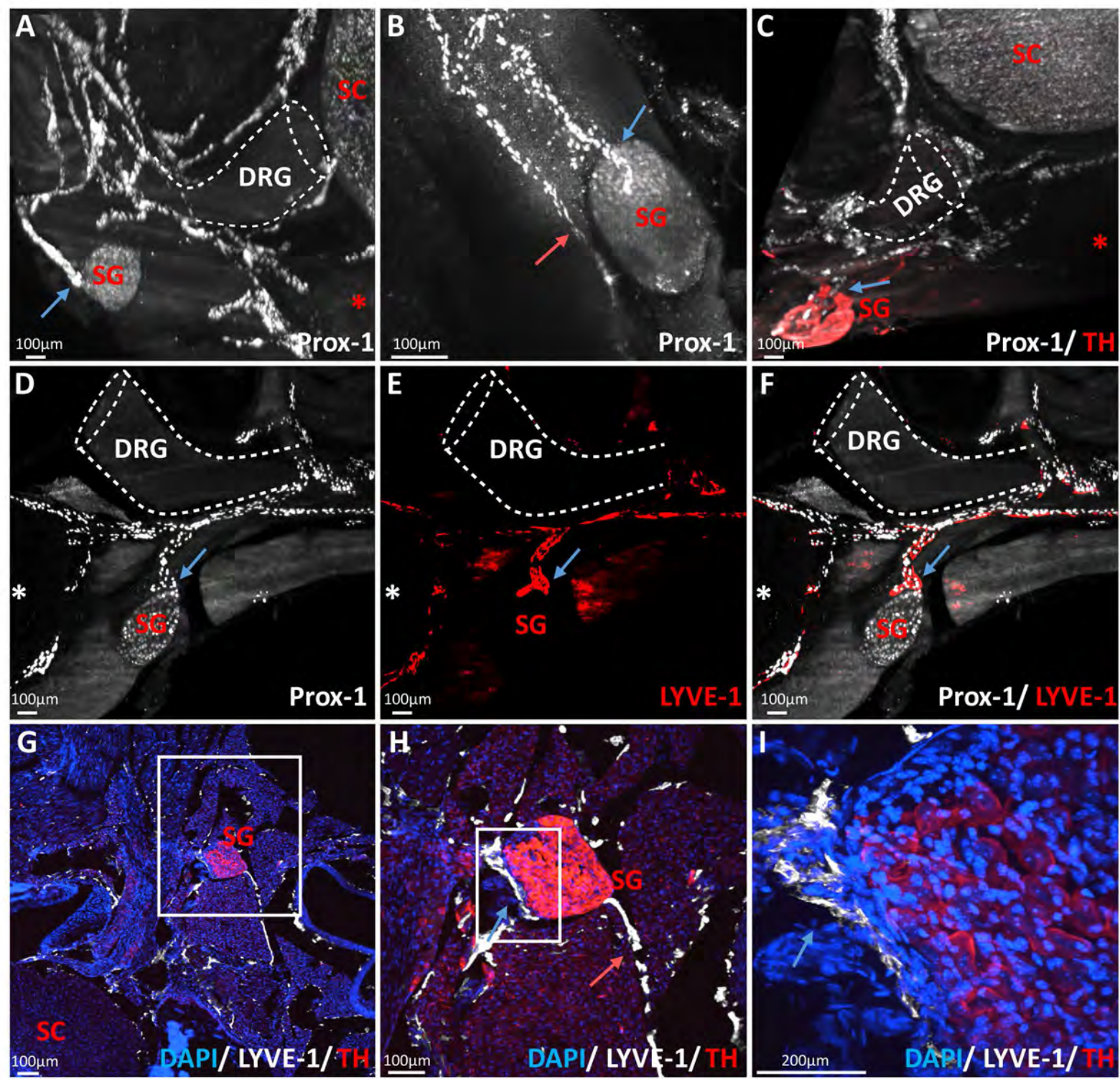

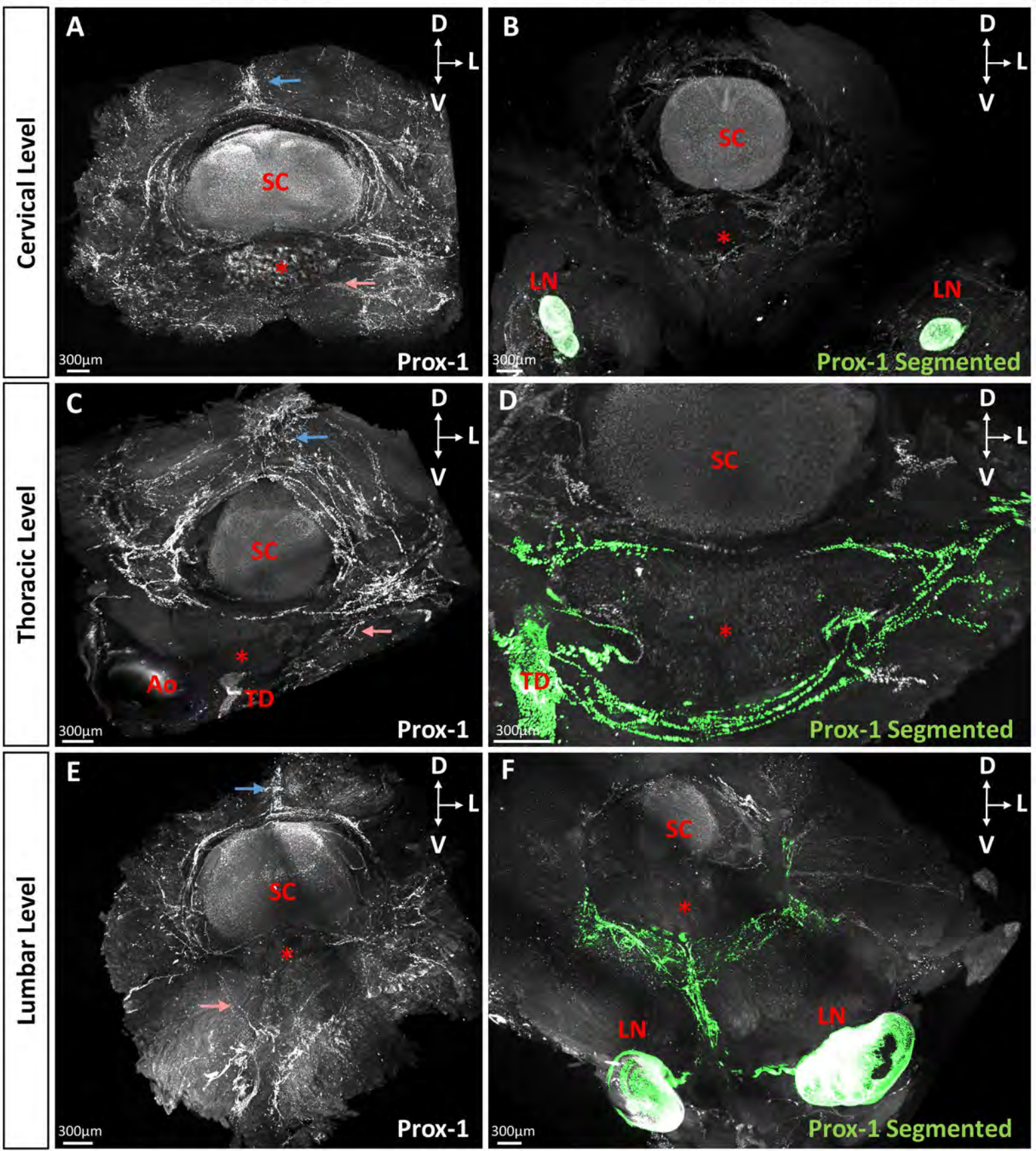

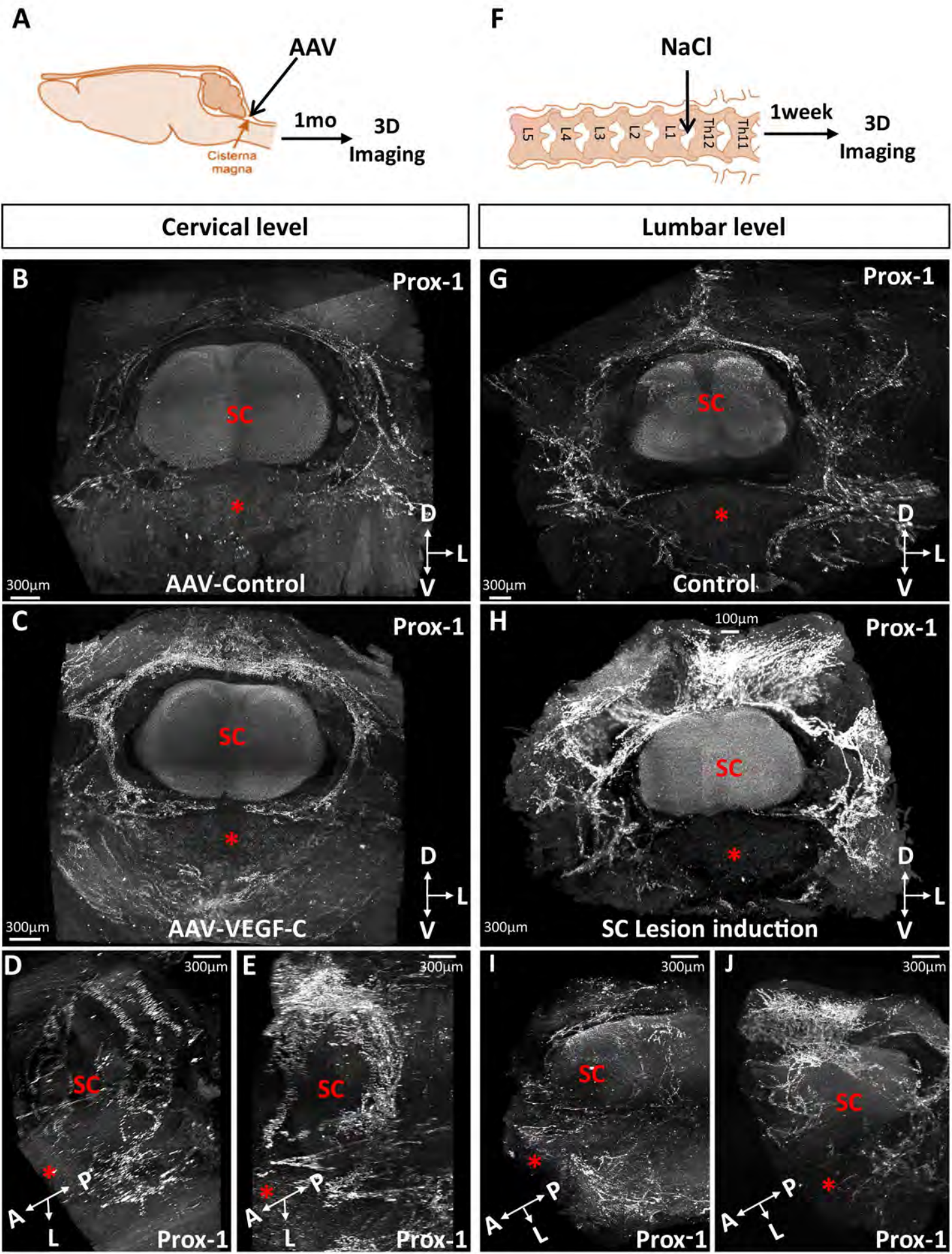

Fig 8

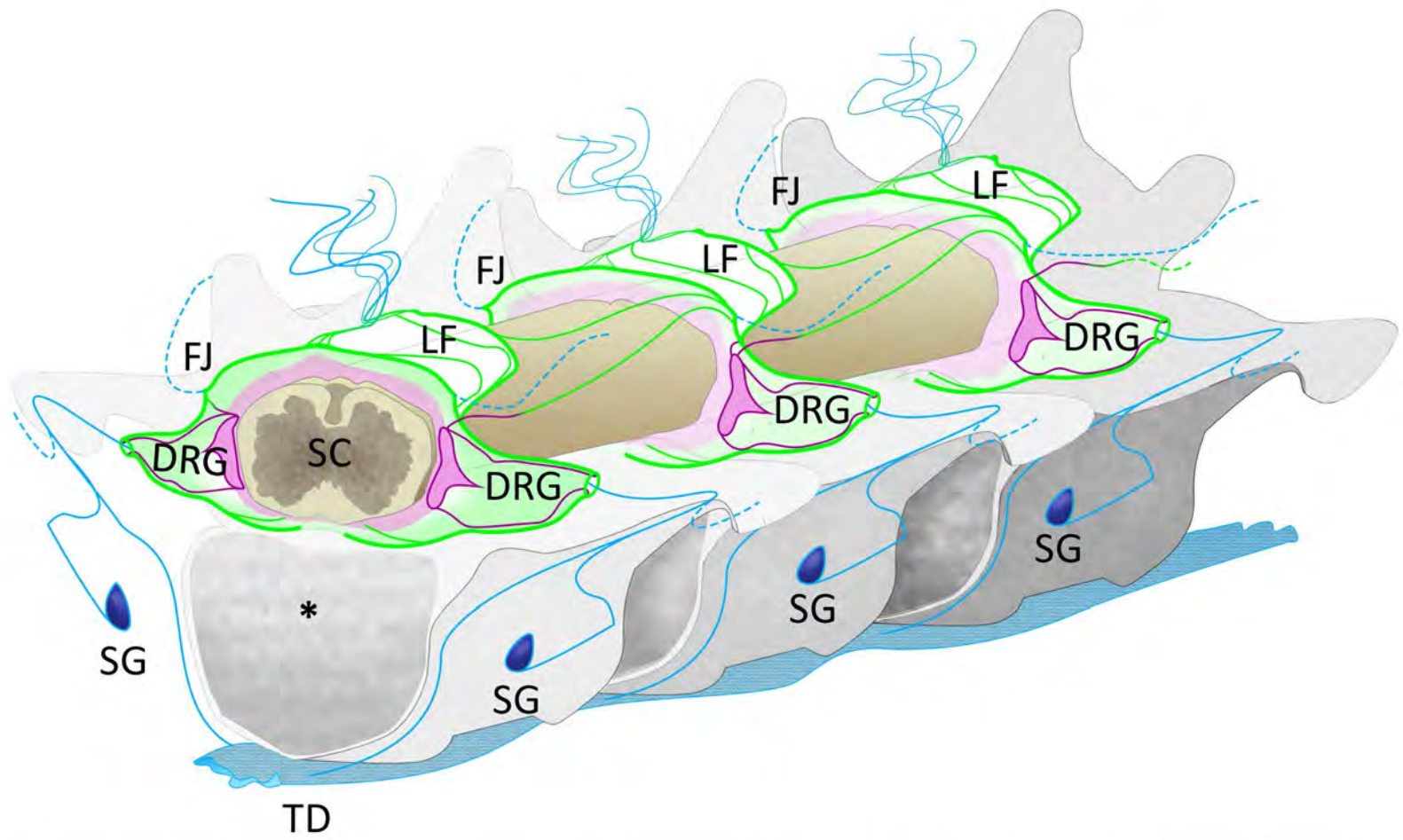

Dural LV networks   Epidural LV networks   Extra-vertebral LV networks

Supp Fig 1

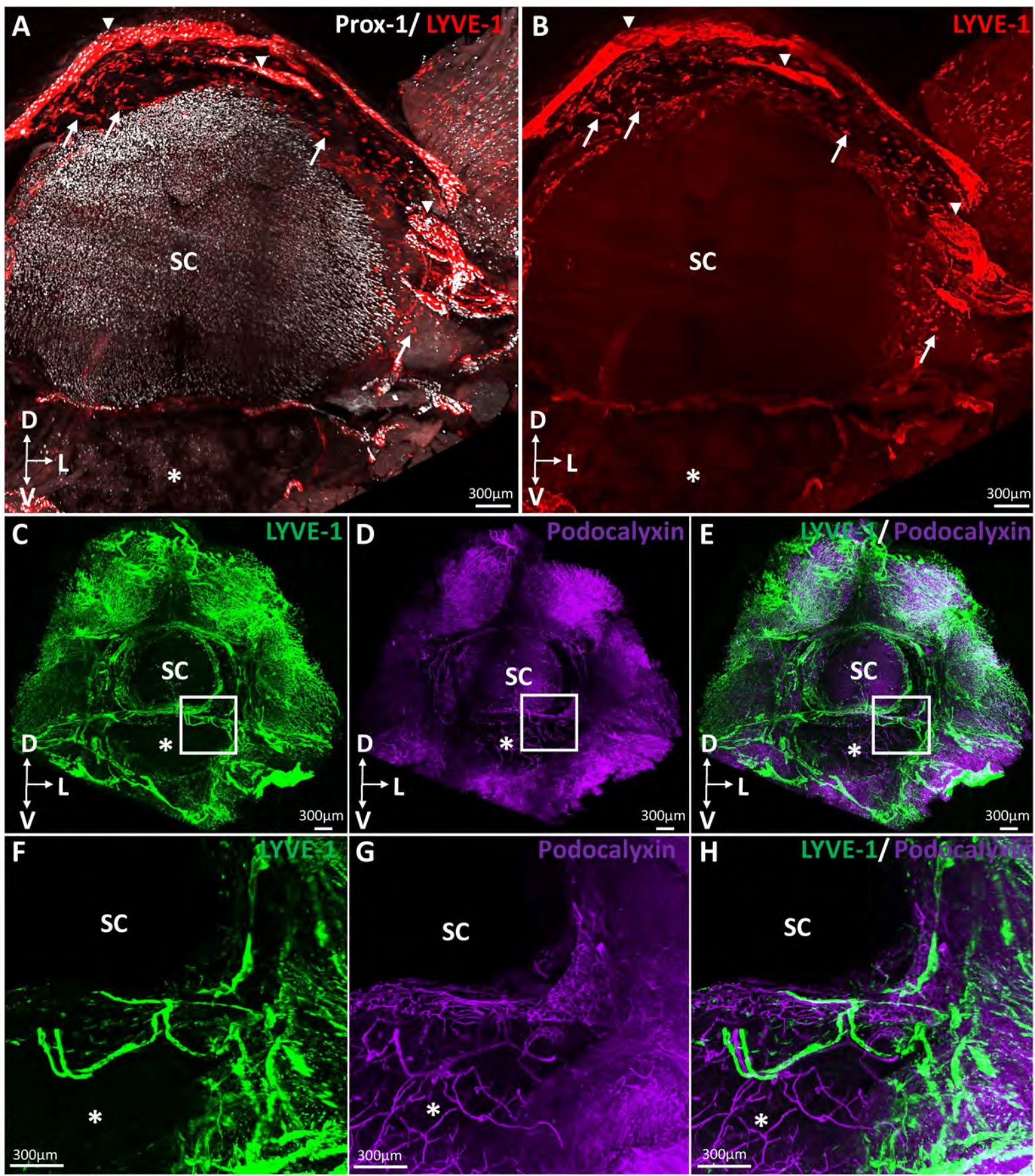

# Supp Fig 2

*Prox1-eGFP* LYVE1

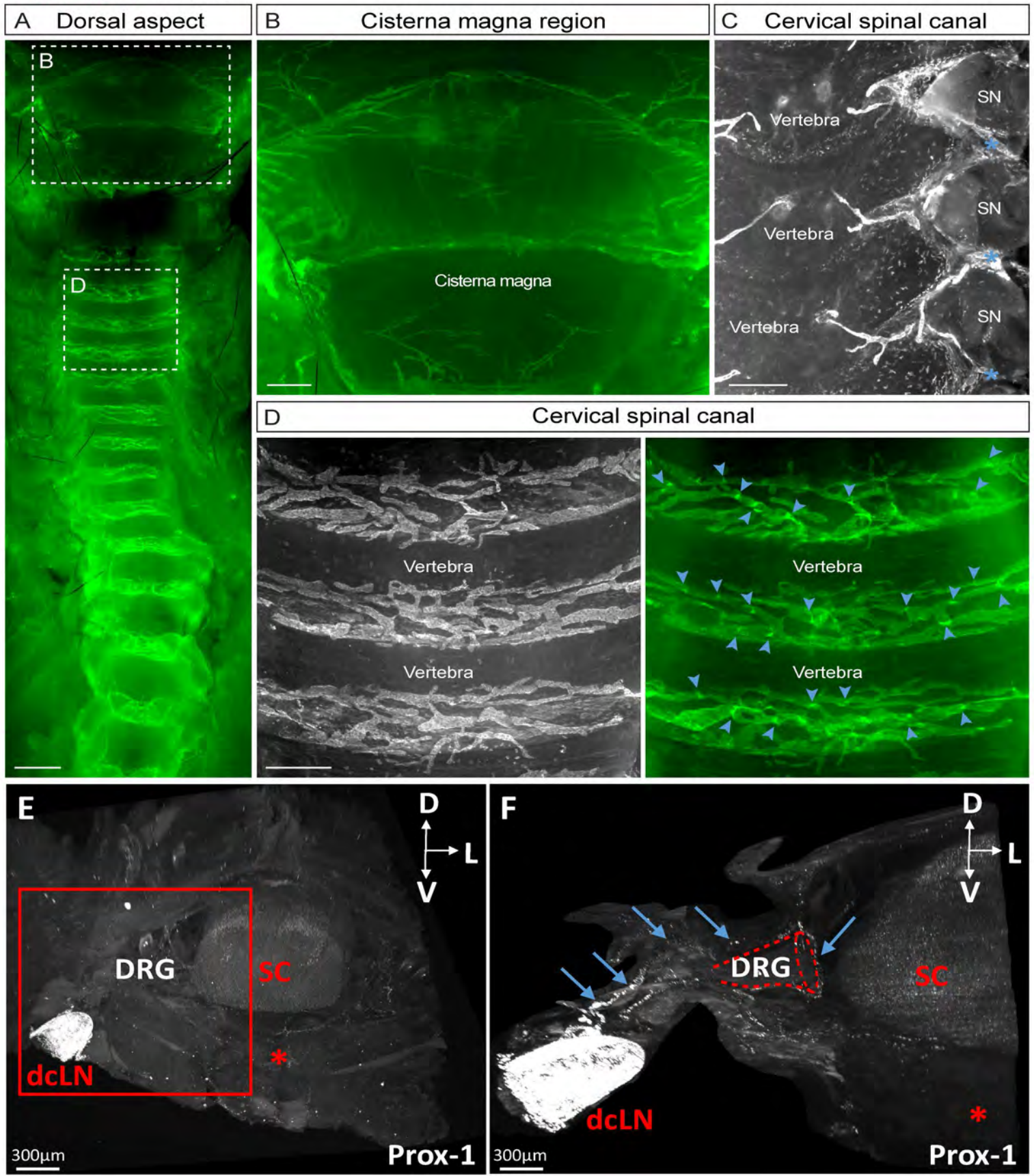